\documentclass[preprint,pra,
nofootinbib]{revtex4}
\usepackage{mathrsfs}
\usepackage{graphicx}
\usepackage{epsfig}
\usepackage{dcolumn}
\usepackage{bm}
\usepackage{amsmath,amssymb,amsthm}
\usepackage[colorlinks=true,linkcolor=blue]{hyperref}
\usepackage{subfigure}
\usepackage{booktabs}
\usepackage[mathscr]{euscript}
\usepackage{ulem}

\newcommand{\bea}{\begin{eqnarray}}
\newcommand{\eea}{\end{eqnarray}}
\newcommand{\beq}{\begin{equation}}
\newcommand{\eeq}{\end{equation}}

\def\/{\over}

\begin{document}

\title{Multi-place nonlocal systems}
\author{S. Y. Lou$^{1,2}$}
\affiliation{
$^{1}$\footnotesize{Center for Nonlinear Science and Faculty of Physical science and technology, Ningbo University, Ningbo, 315211, China}\\
$^2$\footnotesize Shanghai Key Laboratory of Trustworthy Computing, East China Normal University, Shanghai 200062, China }

\begin{abstract}
Two-place nonlocal systems have attracted many scientists' attentions.
In this paper, two-place non-localities are extended to multi-place non-localities. Especially, various two-place and four-place nonlocal nonlinear Schr\"odinger (NLS) systems and Kadomtsev-Petviashvili (KP) equations are systematically obtained from the discrete symmetry reductions of the coupled local systems. The Lax pairs for the two-place and four-place nonlocal NLS and KP equations are explicitly given. Some types of exact solutions especially the multiple soliton solutions for two-place and four-place KP equations are investigated by means of the group symmetric-antisymmetric separation approach.
\end{abstract}

\maketitle
\section{Introduction}
In 2013, Ablowitz and Musslimani proposed a first integrable nonlocal nonlinear model, the nonlinear Schr\"odinger (NLS) equation\cite{AM}
\begin{equation}
iA_{t}+A_{xx}\pm A^2B=0,
\ \ \  B=\hat{f}A=\hat{P}\hat{C}A=A^*(-x,t), \label{Eq1}
\end{equation}
where the operators $\hat{P}$ and $\hat{C}$ are the usual parity and charge conjugation.
In literature, the nonlocal nonlinear Schr\"odinger equation \eqref{Eq1} is also called parity-time reversal ($\hat{P}\hat{T}$) symmetric (more precisely, should be called $\hat{P}\hat{C}$ symmetric).  $\hat{P}$-$\hat{C}$-$\hat{T}$ symmetries play important roles in the quantum physics \cite{2} and many other areas of physics, such as the quantum chromodynamics \cite{3}, electric circuits \cite{4}, optics \cite{5,6}, Bose-Einstein condensates \cite{7}, and so on.

Notice that the model equation \eqref{Eq1} includes two different places $\{x,\ t\}$ and $\{x'=-x,\ t'=t\}$, thus, we call all the models including two places $\{x,\ t\}$ and $\{x',\ t'\}$ two-place systems or Alice-Bob systems \cite{ABs}. Two-place systems may be developed to describe various two-place physics which is the physical theory to explain the corelated/entangled natural phenomena happened at two different places\cite{AB}.

In addition to the nonlocal NLS system \eqref{Eq1}, there are many other types of two-place nonlocal models, such as the nonlocal KdV systems \cite{9,AB,ABs}, nonlocal MKdV systems \cite{10,11,ABs}, nonlocal discrete NLS systems \cite{12}, nonlocal coupled NLS systems \cite{13}, nonlocal Davey-Stewartson systems \cite{14,15,16}, generalized nonlocal NLS equation \cite{TL}, nonlocal nonautonomous KdV equation \cite{Tang1}, nonlocal peakon systems \cite{ABP}, nonlocal KP systems, nonlocal sine Gordon systems and nonlocal Toda systems \cite{AB,ABs}.

In natural sciences, more than two events occurred at different places may be correlated or entangled. To describe multi-place problems, it is natural and important to establish some possible multi-place nonlocal models.

 In section II, we propose two general methods to find a model with multi-place non-localities. In section III, we focus on multi-place nonlocal integrable systems, especially for the two-place and four-place nonlocal NLS equations and KP equations. Section IV is devoted to investigating special solutions of two special two-place and four-place KP systems. The last section is a short summary and discussion.

 \section{Generalized aspect to find multi-place nonlocal systems}
To find multi-place nonlocal systems, there may be several possible approaches. In this section we focus on two simplest methods.
The first method is to find a possible discrete symmetry group with $n$ elements for an $m$ component coupled system such that the discrete symmetry reductions can be found.
The second one is to apply the so-called consistent correlated bang (CCB) for a lower component system to get a higher component system so that the first method can be used.
\subsection{Multi-place nonlocal systems from multi-component systems}
For the $m$-component system
\begin{equation}
K_i(u_1,\ u_2,\ \ldots,\ u_m)=0,\ i=1,\ 2,\ \ldots,\ m,\ \label{Ki}
\end{equation}
where $K_i,\ i=1,\ 2,\ \ldots,\ m$ are functions of $u_j,\ j=1,\ 2,\ \ldots,\ m$ and their derivatives with respect to the space and time variables ${\cal{X}}=\{x_1,\ x_2,\ \ldots,\ x_d,\ t\}$, if we can find an $n$-order discrete group
\begin{equation}
{\cal{G}}=\{\hat{g}_0=I=\mbox{identity},\ \hat{g}_1,\ \ldots,\ \hat{g}_{n-1}\}, \label{G}
\end{equation}
then one may find a suitable transformation
\begin{equation}
u_i=U_i(v_1,\ v_2,\ \ldots, \ v_m),\  i=1,\ 2,\ \ldots,\ m, \label{Ui}
\end{equation}
which transforms the original equation system \eqref{Ki} to a new one
\begin{equation}
\tilde{K}_i(v_1,\ v_2,\ \ldots,\ v_m)=0,\ i=1,\ 2,\ \ldots,\ m,\ \label{K1i}
\end{equation}
thereafter, the $\cal{G}$-symmetry reductions can be directly obtained with some $v_i,\ i=1,\ 2,\ \ldots,\ m$ related to others by suitable group elements $\hat{g},\ j=1,\ 2,\ \ldots,\ n$. Usually, the $\cal{G}$-symmetry reductions are multi-place nonlocal systems if $\hat{g}{\cal{X}}\neq {\cal{X}}$ for some $j$.

Here is a simple special example.
It is clear that the following integrable coupling KP system
\begin{eqnarray} \left\{ \begin{array}{l}
(u_{t}+6uu_x+u_{xxx})_x+\sigma^2u_{yy}=0,\\
{[v_t+6(vu)_x+v_{xxx}]}_x+\sigma^2v_{yy}=0,\\
{[w_t+6(wu)_x+w_{xxx}]}_x+\sigma^2w_{yy}=0,\\
{[z_t+6(zu)_x+6(vw)_x+z_{xxx}]}_x+\sigma^2z_{yy}=0,
\end{array}\right. \label{uvwz}
\end{eqnarray}
possesses an eighth order discrete symmetry group
\begin{equation}
{\cal{G}}={\cal{G}}_1 \cup \hat{C}{\cal{G}}_1,\quad  {\cal{G}}_1\equiv \{1,\ \hat{P}^x\hat{T},\ \hat{P}^y,\ \hat{P}^x\hat{T}\hat{P}^y\}, \label{G1}
\end{equation}
where
the operators $\hat{P}^x,\ \hat{P}^y$, $\hat{T}$ and $\hat{C}$ are the parity for the space variables $x$ and $y$, time reversal and charge conjugate (complex conjugate in mathematics) defined by
\begin{eqnarray}
&&\hat{P}^x x=-x,\quad \hat{P}^y y=-y,\quad \hat{T}t=-t,\quad \hat{C}u=u^*,
\end{eqnarray}
respectively.

Using the symmetry group ${\cal{G}}$, one can directly obtain the following eight discrete symmetry reductions
\begin{eqnarray}
&&p_{xt}+\left\{p_{xx}+6ad(p+p^{\hat{g}})[2d(r-r^{\hat{g}})
+p-p^{\hat{g}}]
+6pu\right\}_{xx}+3\sigma^2p_{yy}=0,\nonumber\\
&&r_{xt}+\left\{r_{xx}-3a(p+p^{\hat{g}})[2d(r-r^{\hat{g}})
+p-p^{\hat{g}}]
-\frac32u(u-4r)\right\}_{xx}+3\sigma^2r_{yy}=0, \label{uvvz}\\
&&p^{\hat{g}}\equiv \hat{g}p,\quad \hat{g}\in {\cal{G}},\ j=0,\ 1,\ \ldots,\ 7,\nonumber
\end{eqnarray}
where $p$ and $r$ are related to $u,\ v,\ w$ and $z$ by the symmetry reduction transformation
\begin{eqnarray}
&&u=p+p^{\hat{g}}+r+r^{\hat{g}},\ v=\frac{a}b(c_0-c_1d+c_2d)(p+p^{\hat{g}}),\nonumber\\
&&w=b(p^{\hat{g}}-p)-2bd(r-r^{\hat{g}}),\ z =c_1p+c_2p^{\hat{g}}+ c_0(r-r^{\hat{g}}). \label{pqrsa}
\end{eqnarray}

For $\hat{g}= \{1,\ \hat{C}\}$, the reductions \eqref{uvvz} are two local integrable coupled KP systems. For $\hat{g}\neq \{1,\ \hat{C}\}$, the reductions \eqref{uvvz} are integrable coupled two-place nonlocal KP systems.

\subsection{Multi-place nonlocal systems from single-component systems via CCB}
To find nonlocal multi-place systems, we can also use the so-called CCB approach proposed in \cite{CCB} from lower-component systems, say, single-component systems. There are three basic steps for the CCB approach, (I) banging a single component equation to a multi-component system, (II) making the banged components correlated, and (III) requiring the correlations are consistent.

For simplicity, we just take the KP equation
\begin{equation}
(u_{t}+6uu_x+u_{xxx})_x+\sigma^2u_{yy}=0\label{KP}
\end{equation}
as a simple example to show the CCB approach.

(I) \bf Bang. \rm To bang the single component KP equation to an $m$-component coupled KP system,
one can make a transformation $u=F(u_0,\ u_1,\ u_2,\ \ldots,\ u_{m-1})$, say,
\begin{equation}
u=\sum_{i=0}^{m-1}u_i.\label{ui}
\end{equation}
Substituting \eqref{ui} into \eqref{KP}, we have
\begin{equation}
\sum_{i=0}^{m-1}\left[\left(u_{it}+6u_i\sum_{j=0}^{m-1}u_{jx}
+u_{ixxx}\right)_x+\sigma^2u_{iyy}\right]=0.\label{KPm}
\end{equation}
It is clear that \eqref{KPm} can be banged to an $m$ component coupled KP system
\begin{equation}
\left(u_{it}+6u_{ix}\sum_{j=0}^{m-1} u_j
+u_{ixxx}\right)_x+\sigma^2u_{iyy}+G_i=0,\ i=0,\ 1,\ 2,\ \ldots,\ m-1,\label{KPmi}
\end{equation}
with $m$ arbitrary functionals $G_i$ under only one condition
\begin{equation}
\sum_{i=0}^{m-1} G_i=0.\ \label{Gi}
\end{equation}

(II) \bf Correlation. \rm To get some nontrivial models, we assume that the banged fields $u_i$ are correlated each other, say, we can write the correlation relations as
\begin{equation}
u_j=\hat{g}u_0,\ j= 0,\ 1,\ 2,\ \ldots,\ m-1.\ \label{corr}
\end{equation}

(III) \bf Consistency. \rm It is natural that the correlation \eqref{corr} and the banged system \eqref{KPmi} should be consistent. Applying $g$ on \eqref{KPmi} for all $i$ and $j$, it is straightforward to prove that the set of the correlated operators $\hat{g}$
\begin{equation}
{\cal{G}}=\{\hat{g}_0,\ \hat{g}_1,\ \hat{g}_2,\ \ldots,\ \hat{g}_{m-1}\} \label{Group}
\end{equation}
consists of an $m$ order finite group. Furthermore, the condition \eqref{Gi} becomes
\begin{equation}
\sum_{i=0}^{m-1} \hat{g}_iG_0=0.\ \label{Gi1}
\end{equation}

It is clear that if take the discrete symmetry group as shown in \eqref{G1} for $m=8$, then we get a four-place nonlocal complex KP equation ($u_0\equiv p$)
\begin{eqnarray}
&&\left(p_{t}+6p_{x}\sum_{j=0}^{7} p^{\hat{g}}
+p_{xxx}\right)_x+\sigma^2p_{yy}+G_0=0\label{KP0}\\
&&\hat{g}\in {\cal{G}}=\left\{1,\ \hat{C},\ \hat{P}^x\hat{T}, \ \hat{P}^y,\ \hat{C}\hat{P}^x\hat{T}, \ \hat{C}\hat{P}^y,\ \hat{P}^x\hat{T}\hat{P}^y,\ \hat{C}\hat{P}^x\hat{T}\hat{P}^y\right\},
\end{eqnarray}
with $G_0$ being a solution of \eqref{Gi1} including $G_0=0$ as a special trivial example.

\section{Two-place and four-place nonlocal integrable systems}
In this section, we apply the general theory of the last section to obtain some multi-place nonlocal extensions for several important physical models such as the NLS, KP, KdV and sine-Gordon models.
\subsection{Two-place and four-place nonlocal NLS systems}
One of the most famous NLS equation
\begin{equation}
iq_{t}+q_{xx}+2\sigma |q|^2q=0,\ \sigma=\pm1, \label{NLS}
\end{equation}
which is a simple reduction of the AKNS system
\begin{eqnarray}
&&iq_{t}+q_{xx}+2\sigma q^2r=0,\\
&&-ir_{t}+r_{xx}+2\sigma r^2q=0, \label{AKNS}
\end{eqnarray}
by using the reduction relation $r= q^*$.

In fact, the AKNS system \eqref{AKNS} possesses a sixteenth order discrete symmetry group
\begin{eqnarray}
&&{\cal{G}}_{\rm AKNS}={\cal{S}}_1 \cup {\cal{S}}_2,\label{GAKNS}\\
&&{\cal{S}}_1  =\hat{E}_{q,r}\{\hat{C},\ \hat{F}\hat{C},\ \hat{C}\hat{P},\ \hat{F}\hat{C}\hat{P},\
\hat{T},\ \hat{F}\hat{T},\ \hat{T}\hat{P},\ \hat{F}\hat{T}\hat{P}\},\label{GAKNS1}\\
&&{\cal{S}}_2 =\hat{C}\hat{E}_{q,r}{\cal{S}}_1 =\{1,\ \hat{F},\ \hat{P},\ \hat{F}\hat{P},\
\hat{C}\hat{T},\ \hat{F}\hat{C}\hat{T},\ \hat{C}\hat{T}\hat{P},\ \hat{F}\hat{C}\hat{T}\hat{P}\},\label{GAKNS2}
\end{eqnarray}
with four second order generators, $\{\hat{F},\ \hat{P},\ \hat{C}\hat{E}_{q,r},\ \hat{T}\hat{E}_{q,r}\}$, where $\hat{P}$ is the shifted parity, $\hat{T}$ is the delayed time reversal, $\hat{C}$ is the charge conjugate, $\hat{F}$ is the field reflection and $\hat{E}_{q,r}$ is the exchange of the fields $q$ and $r$. $\hat{F}$ and $\hat{E}_{q,r}$ are defined by
\begin{eqnarray}
&&\hat{F}\left(\begin{array}{c} q\\ r\end{array}\right)=\left(\begin{array}{c} -q\\ -r\end{array}\right),\
\hat{E}_{q,r}\left(\begin{array}{c} q\\ r\end{array}\right)=\left(\begin{array}{c} r\\ q\end{array}\right). \label{FE}
\end{eqnarray}

From the definition \eqref{FE}, we know that there are two types of discrete symmetries. The first type of symmetries (${\cal{S}}_1$) exchanges the fields $q$ and $r$. However, the second type of symmetries (${\cal{S}}_2$) does not exchange the field variables, and thus it can not be used to obtain nontrivial reductions.
Consequently, the AKNS system \eqref{AKNS} possesses the following eight nontrivial discrete symmetry reductions
\begin{eqnarray}
&&iq_{t}+q_{xx}+2\sigma q^2q^{\hat{g}}=0,\label{AKNS8}\\
&&\hat{g}\in =
\left\{\hat{C},\ \hat{F}\hat{C},\ \hat{C}\hat{P},\ \hat{F}\hat{C}\hat{P},\
\hat{T},\ \hat{F}\hat{T},\ \hat{T}\hat{P},\ \hat{F}\hat{T}\hat{P}  \right\}.\nonumber
\end{eqnarray}
Obviously, the reductions \eqref{AKNS8} include two local reductions for $\hat{g}=\{C,\ FC\}$ and six two-place nonlocal reductions for $\hat{g}\neq \{\hat{C},\ \hat{F}\hat{C}\}$.

To get four-place NLS type nonlocal systems, one has to study the discrete symmetry reductions for some higher component AKNS systems. Here are two special four component AKNS systems
 \begin{eqnarray} \left\{\begin{array}{l}
\mbox{\rm i}q_t+q_{xx}+\frac12\sigma (p+q)[2qr+s(q-p)]=0,\\
\mbox{\rm i}p_t+p_{xx}+\frac12\sigma (p+q)[2ps+r(p-q)]=0,\\
\mbox{\rm i}r_t-r_{xx}-\frac12\sigma (s+r)[2qr+p(r-s)]=0,\\
\mbox{\rm i}s_t-s_{xx}-\frac12\sigma (s+r)[2ps+q(s-r)]=0,
\end{array}
\right.
 \label{CAKNS}
 \end{eqnarray}
 and
\begin{eqnarray}
\left\{\begin{array}{l}
\mbox{\rm i}q_t+q_{xx}+2\sigma (qr+ps)q=0,\\
\mbox{\rm i}p_t+p_{xx}+2\sigma (qr+ps)p=0,\\
\mbox{\rm i}r_t-r_{xx}-2\sigma (qr+ps)r=0,\\
\mbox{\rm i}s_t-s_{xx}-2\sigma (qr+ps)s=0.
\end{array}
\right.
 \label{CAKNSa}
 \end{eqnarray}

It is clear that the coupled AKNS systems \eqref{CAKNS} and \eqref{CAKNSa} will be reduced back to the standard AKNS \eqref{NLS} if $p=q$ and $s=r$.

It is straightforward to find that the coupled AKNS systems \eqref{CAKNS} and \eqref{CAKNSa} possess an common sixteenth order discrete symmetry group
\begin{eqnarray}
{\cal{G}}_{\rm CAKNS}={\cal{G}}_1 \cup \hat{E}_{pq}^{rs}{\cal{G}}_1 \cup \hat{E}_{qs}^{pr}{\cal{G}}_2  \cup \hat{E}_{qr}^{ps}{\cal{G}}_2,\quad {\cal{G}}_1=
\{1,\ \hat{C}\hat{T},\ \hat{P},\ \hat{P}\hat{C}\hat{T} \}, \quad {\cal{G}}_2=\hat{C}{\cal{G}}_1,\label{GCAKNS}
\end{eqnarray}
where the field exchange operators $\hat{E}_{pq}^{rs}$, $\hat{E}_{qs}^{pr}$ and $\hat{E}_{qr}^{ps}$ are defined by
\begin{eqnarray}
\hat{E}_{pq}^{rs}\left(\begin{array}{c}p\\ q\\ r\\ s\end{array}\right)=\left(\begin{array}{c}q\\ p\\ s\\ r\end{array}\right),\ \hat{E}_{qs}^{pr}\left(\begin{array}{c}p\\ q\\ r\\ s\end{array}\right)=\left(\begin{array}{c}r\\ s\\ p\\ q\end{array}\right),\ \hat{E}_{qr}^{ps}\left(\begin{array}{c}p\\ q\\ r\\ s\end{array}\right)=\left(\begin{array}{c}s\\ r\\ q\\ p\end{array}\right)=\hat{E}_{pq}^{rs}\hat{E}_{qs}^{pr}\left(\begin{array}{c}p\\ q\\ r\\ s\end{array}\right).
\end{eqnarray}

In the discrete symmetry group \eqref{GCAKNS}, we have not considered the field reflection operator $\hat{F}$, because the sign change of the fields has been included in the model parameter $\sigma$.

Four types of nontrivial and nonequivalent local or nonlocal AKNS systems can be obtained from the reductions of the discrete symmetry group \eqref{GCAKNS}.

The first type of reductions can be written from \eqref{CAKNS} as
\begin{eqnarray}
&&\left\{\begin{array}{l}
\mbox{\rm i}q_t+q_{xx}+\frac12\sigma (q^{\hat{f}}+q)[2qr+r^{\hat{f}}(q-q^{\hat{f}})]=0,\\
\mbox{\rm i}r_t-r_{xx}-\frac12\sigma (r^{\hat{f}}+r)[2qr+q^{\hat{f}}(r-r^{\hat{f}})]=0,
\end{array}
\right.
 \label{ABAKNS2}\\
&&\hat{f} \in {\cal{G}}_1=\{1,\ \hat{T}\hat{C},\ \hat{P},\ \hat{P}\hat{T}\hat{C}\}, \quad(p,\ s)=\hat{f}(q,\ r).
 \label{ps}
 \end{eqnarray}
 The reduction \eqref{ABAKNS2} is local for $\hat{f}=1$, while the other three reductions of \eqref{ABAKNS2} with $\hat{f}\neq 1$ are two-place nonlocal AKNS systems.

 The second type of AKNS systems obtained
 from \eqref{CAKNSa} reads
 \begin{eqnarray}
&&\left\{\begin{array}{l}
\mbox{\rm i}q_t+q_{xx}+2\sigma q\left(qr+q^{\hat{f}}r^{\hat{f}}\right)=0,\\
\mbox{\rm i}r_t-r_{xx}-2\sigma r\left(qr+q^{\hat{f}}r^{\hat{f}}\right)=0,
\end{array}
\right.
 \label{ABAKNS2a}\\
&&\hat{f} \in {\cal{G}}_1=\{1,\ \hat{T}\hat{C},\ \hat{P},\ \hat{P}\hat{T}\hat{C}\},\quad (p,\ s)=\hat{f}(q,\ r).
 \label{psa}
 \end{eqnarray}
 As in the first type of reductions \eqref{ABAKNS2}, the reduction \eqref{ABAKNS2a} with $\hat{f}=1$ is the local AKNS while the others are two-place nonlocal AKNS systems.

The third type of discrete symmetry reductions from \eqref{CAKNS} possesses the forms
\begin{eqnarray}
&&\left\{\begin{array}{l}
\mbox{\rm i}q_t+q_{xx}+\frac12\sigma (p+q)[2qq^{g}+p^{g}(q-p)]=0,\\
\mbox{\rm i}p_t+p_{xx}+\frac12\sigma (p+q)[2qq^{g}+p^{g}(q-p)]=0,\\
\end{array}
\right. \label{ABAKNS1}\\
&&\hat{g} \in {\cal{G}}_2=\{\hat{C},\ \hat{T},\ \hat{C}\hat{P},\ \hat{P}\hat{T}\},\quad (r,\ s)=\hat{g}(q,\ p).
\label{rs}
\end{eqnarray}
 In this case, the local AKNS system is related to $\hat{g}=\hat{C}$, while the two-place nonlocal AKNS reductions are corresponding to $\hat{g}\neq\hat{C}$.

The fourth type of discrete symmetry reductions
\begin{eqnarray}
&&\left\{\begin{array}{l}
\mbox{\rm i}q_t+q_{xx}+2\sigma q\left(qq^{\hat{g}}+pp^{\hat{g}}\right)=0,\\
\mbox{\rm i}p_t+p_{xx}+2\sigma p\left(qq^{\hat{g}}+pp^{\hat{g}}\right)=0,\\
\end{array}
\right. \label{ABAKNS1a}\\
&&\hat{g} \in {\cal{G}}_2=\{\hat{C},\ \hat{T},\ \hat{C}\hat{P},\ \hat{P}\hat{T}\},\quad (r,\ s)=\hat{g}(q,\ p),
\label{rsa}
\end{eqnarray}
can be obtained from \eqref{CAKNSa}. When $\hat{g}=\hat{C}$, the reduction \eqref{ABAKNS1a} is just the well known local Manakov system
\begin{eqnarray}
&&\left\{\begin{array}{l}
\mbox{\rm i}q_t+q_{xx}+2\sigma q\left(qq^*+pp^*\right)=0,\\
\mbox{\rm i}p_t+p_{xx}+2\sigma p\left(qq^*+pp^*\right)=0.
\end{array}
\right. \label{Manakov}
\end{eqnarray}
When $\hat{g}=\{\hat{T},\ \hat{C}\hat{P},\ \hat{P}\hat{T}\}$, the reductions of \eqref{ABAKNS1a} are two-place nonlocal Manakov  models.

The integrability of the coupled AKNS system \eqref{CAKNS}, the nonlocal AKNS systems \eqref{ABAKNS2} and \eqref{ABAKNS1} can be guaranteed by the following common Lax pair
\begin{eqnarray}
&&\psi_x=\left(\begin{array}{cccc}
\lambda & \frac12(p+q) & 0 & 0 \\
-\frac{\sigma}2(r+s) & -\lambda &0 &0 \\
0 & \frac12(q-p) & \lambda & \frac12(q+p) \\
\frac{\sigma}2(s-r) & 0 & -\frac{\sigma}2(s+r) & -\lambda
\end{array}\right)\psi,\label{Lx}\\
&&\psi_t=\left(\begin{array}{cccc}
u & \diamondsuit (p+q) & 0 & 0\\
-\sigma \diamondsuit(s+r) & -u & 0 & 0\\
-v & \diamondsuit  (q-p) & u & \diamondsuit (p+q)\\
\diamondsuit (s-r) & v & -\diamondsuit (s+r) & -u
\end{array}\right)\psi,\label{Lt}
\end{eqnarray}
where
$$ u\equiv \frac{\mbox{\rm i}}4[\sigma(s+r)(p+q)+8\lambda^2],\ v\equiv \frac{\mbox{\rm i}}2\sigma(sp-qr),\ \diamondsuit \equiv \frac{\mbox{\rm i}}2(2\lambda+\partial_x).$$

The integrability of the coupled AKNS system \eqref{CAKNSa}, the nonlocal AKNS systems \eqref{ABAKNS2a} and \eqref{ABAKNS1a} can be ensured by the Lax pair of the two component vector AKNS system,
\begin{eqnarray}
&&\psi_x=\left(\begin{array}{ccc}
\lambda_1-\lambda & q & p \\
-\sigma r & -\lambda &0  \\
-\sigma s & 0 & -\lambda
\end{array}\right)\psi,\label{LaxAKNS}\\
&&\psi_t=\mbox{\rm i} \sigma \left(\begin{array}{ccc}
\sigma \lambda_1^2-c\lambda^2+qr+ps & \sigma (q_x+\lambda_1q) & \sigma (p_x+\lambda_1p)\\
r_x-\lambda_1r & -c\lambda^2-qr & -pr\\
s_x-\lambda_1s & -qs & -c\lambda^2 -ps
\end{array}\right)\psi,\label{LatAKNS}
\end{eqnarray}
where $c,\ \lambda$ and $\lambda_1$ are arbitrary constants.

It is interesting that some known integrable nonlocal NLS (or named ABNLS) systems are just the special reductions of the nonlocal AKNS systems \eqref{ABAKNS2}, \eqref{ABAKNS2a} \eqref{ABAKNS1} and \eqref{ABAKNS1a}. For instance, taking $p=q=A,\ s=r=B$ in \eqref{ABAKNS1}, we get the known nonlocal NLS systems \eqref{AKNS8} and some others such as those in \cite{AM,ABs,AM1} and \cite{Zhang},
\begin{eqnarray}
&&
\mbox{\rm i}A_t+A_{xx}+2\sigma A^2B=0, \label{ABNLS1}\\
&&B=\hat{g}A,\quad \hat{g} \in \{\hat{T},\ \hat{C}\hat{P},\ \hat{P}\hat{T}\}.
 \label{BA1}
 \end{eqnarray}
In addition to the known nonlocal NLS reductions \eqref{ABNLS1}, one can also obtain some types of \emph{\textbf{novel}} local and nonlocal two-place and four-place NLS type systems from the AKNS systems \eqref{ABAKNS2}, \eqref{ABAKNS2a}, \eqref{ABAKNS1}  and \eqref{ABAKNS1a}.

It is clear that \eqref{ABAKNS2} allows a special reduction $r=q^*\equiv A^*$ and then
\begin{eqnarray}
&&\mbox{\rm i}A_t+A_{xx}+\frac12\sigma (B+A)[2AA^*+B^*(A-B)]=0,
 \label{ABNLS2}\\
&&B=\hat{f}A,\quad \hat{f} \in \{\hat{P},\ \hat{C}\hat{T},\ \hat{P}\hat{C}\hat{T}\}.
 \label{ps1}
 \end{eqnarray}

In fact, from the coupled AKNS systems \eqref{CAKNS} and \eqref{CAKNSa}, we can get 32 different types of NLS reductions.
Applying the symmetry group $\cal{G}$ to \eqref{CAKNS}, we have
\begin{eqnarray}
&&
\mbox{\rm i}q_t+q_{xx}+\frac12\sigma (q^{\hat{f}}+q)[2qq^{\hat{g}}+q^{\hat{f}\hat{g}}(q-q^{\hat{f}})]=0,\
(p,\ r,\ s)=(q^{\hat{f}},\ q^{\hat{g}},\ q^{\hat{f}\hat{g}}),\ \label{ABNLS3}\\
&&\hat{g} \in {\cal{G}}_1^c=\{\hat{C},\ \hat{T},\ \hat{C}\hat{P},\ \hat{P}\hat{T}\}, \
\hat{f} \in {\cal{G}}_1=\{1,\ \hat{P},\ \hat{C}\hat{T},\ \hat{P}\hat{C}\hat{T}\}.
 \label{prs}
 \end{eqnarray}
The full $\hat{P}$-$\hat{T}$-$\hat{C}$ symmetry reductions of \eqref{CAKNSa} possess the form
\begin{eqnarray}
&&\begin{array}{l}
\mbox{\rm i}q_t+q_{xx}+2\sigma q\left(qq^{\hat{g}}+q^{\hat{f}}q^{\hat{f}\hat{g}}\right)=0,\
(p,\ r,\ s)=(\hat{f}q\equiv q^{\hat{f}},\ \hat{g}q\equiv q^{\hat{g}},\ \hat{f}\hat{g}q\equiv q^{\hat{f}\hat{g}}),
\end{array} \label{ABNLS3a}\\
&&\hat{g} \in {\cal{G}}_1^c=\{\hat{C},\ \hat{T},\ \hat{C}\hat{P},\ \hat{P}\hat{T}\},\ \hat{f} \in {\cal{G}}_1=\{1,\ \hat{C}\hat{T},\ \hat{P},\ \hat{C}\hat{P}\hat{T}\}.
\label{prsa}
\end{eqnarray}

For the sixteen reductions \eqref{ABNLS3}, there are one local case
($\hat{f}=1,\ \hat{g}=\hat{C}$), nine two-place cases \eqref{ABAKNS1} ($\hat{f}=1,\ \hat{g}=\{\hat{T},\ \hat{P}\hat{C},\ \hat{P}\hat{T}\}$), \eqref{ABAKNS2} ($\hat{g}=\hat{C},\ \hat{f}=\{\hat{P},\ \hat{T}\hat{C},\ \hat{P}\hat{T}\hat{C}\}$) and the cases related to $\hat{g}=\hat{C}\hat{f},\ \hat{f}=\{\hat{P},\ \hat{T}\hat{C},\ \hat{P}\hat{T}\hat{C}\}$,
\begin{eqnarray}
&&
\mbox{\rm i}q_t+q_{xx}+\frac12\sigma (p+q)[2qr+s(q-p)]=0, \label{ABNLS4}\\
&&(p,\ r,\ s)=(\hat{f}q,\ \hat{f}q^*,\ q^*),\quad  \hat{f} \in \{\hat{P},\ \hat{C}\hat{T},\ \hat{P}\hat{C}\hat{T}\}.
 \label{prs4}
 \end{eqnarray}
 All other six cases,
 ($\{\hat{f}=\hat{P},\ \hat{g}=(\hat{T},\ \hat{P}\hat{T})\},\ \{\hat{f}=\hat{T}\hat{C},\ \hat{g}=(\hat{C}\hat{P},\ \hat{P}\hat{T})\},\ \{\hat{f}=\hat{P}\hat{T}\hat{C},\ \hat{g}=(\hat{T},\ \hat{P}\hat{C})\}$) are four-place nonlocal NLS equations which have not yet appeared in literature. For instance, for $\hat{g}=\hat{C}\hat{P}$ and $\hat{f}=\hat{C}\hat{T}$, the related four-place nonlocal NLS equation \eqref{ABNLS3} becomes
\begin{eqnarray}
\mbox{\rm i}q_t+q_{xx}+\frac12\sigma (q^*(x,-t)+q)[2qq^*(-x,t)+q(-x,-t)(q-q^*(x,-t))]=0.  \label{ABNLS5}
\end{eqnarray}
The systems \eqref{ABNLS4} and \eqref{ABNLS5} are called four-place nonlocal NLS equation because four places $(x,\ t)$, $(x,\ -t)$, $(-x,\ t)$ and $(-x,\ -t)$ are included.

Similarly, for the sixteen reductions \eqref{ABNLS3a}, there are one local case, nine two-place nonlocal cases and six four-place nonlocal cases,
\begin{eqnarray}
&&
\mbox{\rm i}q_t+q_{xx}+2\sigma q\left(qq^{\hat{g}}+q^{\hat{f}}q^{\hat{f}\hat{g}}\right)=0,\
(p,\ r,\ s)=(\hat{f}q,\ \hat{g}q,\ \hat{f}\hat{g}q),
 \label{ABNLS3a4}\\
&&(\hat{g},\ f)=(\hat{T},\ \hat{P}\{1,\ \hat{C}\hat{T}\}),
(\hat{C}\hat{P},\ \hat{C}\hat{T}\{1,\ \hat{P}\}),\ (\hat{P}\hat{T},\ \{\hat{C}\hat{T},\ \hat{P}\}).
\label{prsa4}
\end{eqnarray}

In fact, there are many other coupled (and decoupled) integrable AKNS systems, say, the vector and matrix AKNS systems. Starting from every coupled (and decoupled) AKNS systems, one may obtain some possible multi-place integrable discrete symmetry reductions.

Here, we just list another two sets of integrable local and nonlocal NLS type systems
\begin{eqnarray}
&&iq_t+q_{xx}+\left[\alpha(qq^{\hat{g}}+q^{\hat{f}} q^{\hat{f}\hat{g}})
+\beta(q^{\hat{f}}q^{\hat{g}}+qq^{\hat{f}\hat{g}}) \right]q=0,\label{KPR}\\
&& \hat{f}\in {\cal{G}}_1,\quad \hat{g}\in {\cal{G}}_2,\nonumber
\end{eqnarray}
and
\begin{eqnarray}
&&iq_t+\alpha (q+q^{f})_{xx}+\gamma (q-q^{f})_{xx}+\left[\beta(q+q^{\hat{f}})^2+ \delta(q-q^{\hat{f}})^2\right]q^{\hat{f}\hat{g}}
\nonumber\\
&&\qquad +\left[\beta(q+q^{\hat{f}})^2-\delta(q-q^{\hat{f}})^2 \right]q^{\hat{g}}=0,\\
&& \hat{f}\in {\cal{G}}_1,\quad \hat{g}\in {\cal{G}}_2,\nonumber
\end{eqnarray}
with free parameters $\alpha,\ \beta,\ \gamma$ and $\delta$, where ${\cal{G}}_1$ and ${\cal{G}}_2$ are given by \eqref{GCAKNS}.

It is clear that when $\beta=0$, the models \eqref{KPR} will be degenerated to \eqref{ABNLS3a}. For convenience, we rewrite \eqref{KPR} as
\begin{eqnarray}
&&iq_t+q_{xx}+V_{\hat{f},\hat{g}} q=0,\label{NLSV}\\
&&V_{\hat{f},\hat{g}}=\alpha(qq^{\hat{g}}
+q^{\hat{f}}q^{\hat{f}\hat{g}})
+\beta(q^{\hat{f}}q^{\hat{g}}+qq^{\hat{f}\hat{g}}),\label{Vfg}
\end{eqnarray}
where $V_{\hat{f},\hat{g}}$ is clearly ${\cal{G}}_{\hat{f},\hat{g}}$ invariant,
\begin{equation} {\cal{G}}_{\hat{f},\hat{g}}V_{\hat{f},\hat{g}}
=V_{\hat{f},\hat{g}},\label{GV}
\end{equation}
\begin{equation}
{\cal{G}}_{\hat{f},\hat{g}}=\{1,\ \hat{f},\ \hat{g},\ \hat{f}\hat{g}
\}. \label{Gfg}
\end{equation}
For concreteness, we list all the independent NLS systems included in \eqref{KPR} (i.e., \eqref{NLSV}) below.\\
(i). ${\cal{G}}_{\hat{C}}\equiv{\cal{G}}_{1,\hat{C}}$ invariant local NLS equation,
\begin{eqnarray}
V_{\hat{f},\hat{g}}=2(\alpha+\beta)qq^*. \label{V0}
\end{eqnarray}
(ii). ${\cal{G}}_{\hat{P}\hat{C}}\equiv {\cal{G}}_{1,\hat{P}\hat{C}}$ invariant two-place nonlocal NLS system,
\begin{eqnarray}
V_{\hat{f},\hat{g}}=2(\alpha+\beta)qq^*(-x,t). \label{V1}
\end{eqnarray}
(iii). ${\cal{G}}_{\hat{P}\hat{T}}\equiv {\cal{G}}_{1,\hat{P}\hat{T}}$ invariant two-place nonlocal NLS system,
\begin{eqnarray}
V_{\hat{f},\hat{g}}=2(\alpha  + \beta)qq(-x,-t). \label{V2}
\end{eqnarray}
(iv). ${\cal{G}}_{\hat{T}}\equiv {\cal{G}}_{1,\hat{T}}$ invariant two-place nonlocal NLS system,
\begin{eqnarray}
V_{\hat{f},\hat{g}}=2(\alpha+\beta)qq(x,-t). \label{V3}
\end{eqnarray}
(v). ${\cal{G}_{\hat{P},\hat{C}}}$ invariant two-place nonlocal NLS system,
\begin{eqnarray}
V_{\hat{f},\hat{g}}=\alpha [qq^*+q(-x,t)q^*(-x,t)]
+ \beta[q(-x,t)q^*+qq^*(-x,t)]. \label{V4}
\end{eqnarray}
(vi). ${\cal{G}_{\hat{P}\hat{T},\hat{C}}}$ invariant two-place nonlocal NLS system,
\begin{eqnarray}
V_{\hat{f},\hat{g}}=\alpha[qq^*+q(-x,-t)q^*(-x,-t)]+ \beta[q^*q^*(-x,-t)+qq(-x,-t)]. \label{V5}
\end{eqnarray}
(vii). ${\cal{G}_{\hat{T},\hat{C}}}$ invariant two-place nonlocal NLS system,
\begin{eqnarray}
V_{\hat{f},\hat{g}}=\alpha[qq^*+q(x,-t)q^*(x,-t)]+ \beta[qq(x,-t)+q^*q^*(x,-t)]. \label{V6}
\end{eqnarray}
(viii). ${\cal{G}_{\hat{P}\hat{T},\hat{P}\hat{C}}}$ invariant four-place nonlocal NLS system,
\begin{eqnarray}
V_{\hat{f},\hat{g}}=\alpha[qq^*(-x,t)+q(-x,-t)q^*(x,-t)]+ \beta[qq(-x,-t)+q^*(-x,t)q^*(x,-t)]. \label{V7}
\end{eqnarray}
(ix). ${\cal{G}_{\hat{P}\hat{C}\hat{T},\hat{T}}}$ invariant four-place nonlocal NLS system,
\begin{eqnarray}
V_{\hat{f},\hat{g}}=\alpha[qq^*(-x,t)+q(-x,-t)q^*(x,-t)]+ \beta[qq(-x,-t)+q^*(-x,t)q^*(x,-t)]. \label{V8}
\end{eqnarray}
(x). ${\cal{G}_{\hat{P},\hat{T}}}$ invariant four-place nonlocal NLS system,
\begin{eqnarray}
V_{\hat{f},\hat{g}}=\alpha[qq(x,-t)+q(-x,t)q(-x,-t)]+ \beta[q(-x,t)q(x,-t)+qq(-x,-t)]. \label{V9}
\end{eqnarray}
Other types of selections of $f$ and $g$ are related to the exchanges of the constants $\alpha$ and $\beta$. All sixteen cases of \eqref{ABNLS3a} can be obtained from the above cases by setting $\beta=0$ or $\alpha=0$.

The first four cases are just known results of the discrete symmetry reductions from the usual AKNS system.

The integrability of \eqref{KPR} (i.e., \eqref{NLSV}) is trivial
because it is only a special discrete symmetry reduction of the so-called (N+M)-component integrable AKNS system (Eqs. (104,105) of \cite{LH97} with $\{\psi,\psi^*,y\}\rightarrow \{q,p,it\}$)
\begin{eqnarray}
&&iq_{kt}=-q_{kxx}+\sum_{n=1}^N\sum_{m=1}^Ma_{nm}q_np_mq_k,\ k=1,\ 2,\ \ldots N,\label{KPq}\\
&&ip_{jt}=p_{jxx}-\sum_{n=1}^N\sum_{m=1}^Ma_{nm}q_np_mp_j,\ j=1,\ 2,\ \ldots M,\label{KPp}
\end{eqnarray}
for $M=N=2$ and special selections of constants $a_{nm}$.
The integrability of \eqref{KPq}-\eqref{KPp} is guaranteed because it is only a symmetry reduction of the KP equation \cite{LH97,LC02}.

It is also interesting to mention that using the $\hat{P}$-$\hat{T}$-$\hat{C}$ symmetry group, one can find more discrete symmetry reductions from all the above reduced model equations. For instance, starting from the well known Manakov systems \eqref{Manakov}, one can find not only the two-place physically significant nonlocal complex systems listed in \cite{Yang}, but also the following two-place and four-place physically significant nonlocal real nonlinear systems, we omit the details on the similar derivation of these reductions
\begin{eqnarray}
&&p_t+p^{\hat{f}}_{xx}+2\sigma p^{\hat{f}}[p^2+(p^{\hat{f}})^2+(p^{\hat{g}})^2 +(p^{\hat{f}\hat{g}})^2]=0,\label{MR}\\
&&\hat{f}\in \{\hat{T},\ \hat{P}\hat{T}\},\ \hat{g}\in \{1,\ \hat{T},\ \hat{P},\ \hat{P}\hat{T}\}.\nonumber
\end{eqnarray}
Especially, if $\hat{g}=1$, two-place models of \eqref{MR}
\begin{eqnarray}
&&p_t+p_{xx}(x,-t)+4\sigma p(x,-t)
\big[p^2+p(x,-t)^2\big]=0,\label{MR21}\\
&&p_t+p_{xx}(-x,-t)+4\sigma p(-x,-t)\big[p^2 +p(-x,-t)^2\big]=0,\label{MR22}
\end{eqnarray}
can also be derived from the usual local NLS equation. There exist only two independent four-place nonlocal systems included in \eqref{MR},
\begin{eqnarray}
&&p_t+p_{xx}(x,-t)+2\sigma p(x,-t)
\big[p^2+p(x,-t)^2+p(-x,t)^2 +p(-x,-t)^2\big]=0,\label{MR41}
\end{eqnarray}
and
\begin{eqnarray}
&&p_t+p_{xx}(-x,-t)+2\sigma p(-x,-t)\big[p^2+p(x,-t)^2+p(-x,t)^2 +p(-x,-t)^2\big]=0.\label{MR42}
\end{eqnarray}

\subsection{Two-place and four-place nonlocal KP systems}
To find multi-place nonlocal KP systems, we have to get some multi-component coupled KP equations.
To guarantee the integrability, we start from the matrix Lax pairs for matrix KP equations
\begin{eqnarray}
&&\psi_{xx}+U\psi+\sigma\psi_y=0,\label{lxkp}\\
&&\psi_t+4\psi_{xxx}+6U\psi_x+3\left(U_x-\int U_y\mbox{\rm dx}\right)\psi=0,\label{ltkp}
\end{eqnarray}
where $\psi$ is an $m$ component vector and $U$ is an $m\times m$ matrix.

The compatibility condition $\psi_{yt}=\psi_{ty}$ of the Lax pair reads
\begin{eqnarray}
&&(U_t+U_{xxx}+3(U_xU+UU_x)+3\sigma [U,\ W])_x+3\sigma^2 U_{yy}=0, \label{MatKP}\\
&&[U,\ W]\equiv UW-WU,\quad W_x=U_y. \label{UW}
\end{eqnarray}
For the non-Abelian complex matrix KP system \eqref{MatKP} with $\sigma={\rm i}=\sqrt{-1}$, its $\hat{P}\hat{T}\hat{C}$ symmetry group is constructed by the generator operators $\hat{P}^x\hat{T}$ and $\hat{C}\hat{P}^y$,
\begin{eqnarray}
{\cal{G}}_n=\left\{1,\ \hat{P}^x\hat{T},\ \hat{C}\hat{P}^y,\ \hat{C}\hat{P}^y\hat{P}^x\hat{T}\right\}. \label{Gn}
\end{eqnarray}
For the Abelian matrix KP system, $[U,\ W]=0$, the $\hat{P}\hat{T}\hat{C}$ symmetry group is the same as given in \eqref{G1} with three generators $\hat{P}^x\hat{T},\ \hat{C}$ and $\hat{P}^y$.

Here, we just list some special examples and the related $\hat{P}\hat{T}\hat{C}$ symmetry reductions.

Example 1. Abelian matrix KP system \eqref{MatKP} with
\begin{eqnarray}
&&U=\left(\begin{array}{cccc}
u& 0& 0& 0\\
w& u& 0& 0 \\
v& 0& u& 0\\
z& v& w & u
\end{array}\right),\label{Uex1}\\
&& u=(1+\hat{f})(1+\hat{g})p,\
v=(1-\hat{f})(1+\hat{g})p,\ w=[(1+\hat{f})(1-\hat{g})p,\  z=(1-\hat{f})(1-\hat{g})p\nonumber
\end{eqnarray}
possesses a single component $\hat{P}\hat{T}\hat{C}$ symmetry reduction
\begin{eqnarray}
&&p_{xt}+\left\{p_{xx}-\frac{3u^2}{4}+6pu
+\frac32[(p-p^{\hat{f}\hat{g}})^2-(p^{\hat{f}} -p^{\hat{g}})^2]\right\}_{xx}+3\sigma^2p_{yy}=0,\label{KPe1}\\
&& \hat{f},\ \hat{g},\in {\cal{G}}_n.\nonumber\\
\end{eqnarray}

Example 2. From the Abelian matrix KP system \eqref{MatKP} with
\begin{eqnarray}
&&U=\left(\begin{array}{cccc}
u& 0& 0& 0\\
 w& u& 0& 0 \\
  v& 0& u& 0\\
   z& 0& 0 & u
\end{array}\right),\label{Uex2}\\
&& u=(1+\hat{f})(1+\hat{g})p,\ v=(1-\hat{f})(1+\hat{g})p,\ w=(1+\hat{f})(1-\hat{g})p,\ z=(1-\hat{f})(1-\hat{g})p,\nonumber
\end{eqnarray}
we can find a $\hat{P}\hat{T}\hat{C}$ symmetry reduction
\begin{eqnarray}
&&p_{xt}+\left\{p_{xx}-\frac{3u^2}{4}+6pu\right\}_{xx} +3\sigma^2p_{yy}=0.\label{KPe2}
\end{eqnarray}

Example 3. From the non-Abelian matrix KP system \eqref{MatKP} with
\begin{eqnarray}
&&U=\left(\begin{array}{cc}
p+2q-r& q-2r+s\\
p-2q+r& s+2r-q
\end{array}\right),\label{Uex3}
\end{eqnarray}
we can find a $\hat{f}\hat{g}$ symmetry reduction
\begin{eqnarray}
&&p_{xt}+3\sigma^2p_{yy}+3\sigma [(2q_1-r_1)p-(2q-r)p_1+s(p_1-2q_1+r_1)-(p-2q+r)s_1]_x \nonumber\\
&&\quad +[p_{xx}+3(p-2q+r)s+3p(p+2q-r)]_{xx}=0,\label{NA}\\
&&(p_1,\ q_1,\ r_1, \ s_1)_x =(p,\ q,\ r, \ s)_y, \\
&& q=p^{\hat{f}},\ r=p^{\hat{g}},\ s=p^{\hat{f}\hat{g}},\ \hat{f}^2=\hat{g}^2=1,\label{KPI}
\end{eqnarray}
where
\begin{equation}
\hat{f},\ \hat{g} \in {\cal{G}}_I=\{1,\ \hat{P}^x\hat{T},\ \hat{C}\hat{P}^y,\ \hat{P}^x\hat{T}\hat{C}\hat{P}^y\}\label{GI}
\end{equation}
for KPI system ($\sigma=\mbox{\rm i}=\sqrt{-1}$) and
\begin{equation}
\hat{f},\ \hat{g} \in {\cal{G}}_{II}=\{1,\ \hat{P}^x\hat{T},\ \hat{C},\ \hat{P}^x\hat{T}\hat{C}\}\label{GII}
\end{equation}
for KPII system ($\sigma=1$).

For the KPI case, the reduction \eqref{NA} contains one usual local KPI reduction,
\begin{eqnarray}
&&p_{xt}-3p_{yy}+(p_{xx}+6p^2)_{xx}=0,\hat{f}=\hat{g},\label{f=g}
\end{eqnarray}
six two-place nonlocal Abel KPI reductions
\begin{eqnarray}
&&p_{xt}-3p_{yy}+[p_{xx}+6p^2 +3(p-p^{\hat{g}})^2]_{xx}=0,\label{f=1}\\
&&\hat{f}=1,\ \hat{g}\in \{\hat{P}^x\hat{T},\ \hat{P}^y\hat{C},\ \hat{P}^x\hat{T}\hat{P}^y\hat{C}\},\nonumber
\end{eqnarray}
and
\begin{eqnarray}
&&p_{xt}-3p_{yy}+[p_{xx}+6p^{\hat{f}} (2p-p^{\hat{f}})]_{xx}=0,\label{g=1}\\
&&\hat{g}=1,\ \hat{f}\in \{\hat{P}^x\hat{T},\ \hat{P}^y\hat{C},\ \hat{P}^x\hat{T}\hat{P}^y\hat{C}\},\nonumber
\end{eqnarray}
and six four-place non-Abelian nonlocal systems,
\begin{eqnarray}
&&p_{xt}-3p_{yy}+3\mbox{\rm i} [(2p^{\hat{f}}_1-p^{\hat{g}}_1)p-(2p^{\hat{f}}-p^{\hat{g}})p_1 +s(p_1-2p^{\hat{f}}_1+p^{\hat{g}}_1) -(p-2p^{\hat{f}}+p^{\hat{g}})p^{\hat{f}\hat{g}}_1]_x \nonumber\\
&&\quad +[p_{xx}+3(p-2p^{\hat{f}}+p^{\hat{g}})p^{\hat{f}\hat{g}} +3p(p+2p^{\hat{f}}-p^{\hat{g}})]_{xx}=0,\  p_{1x} =p_y, \label{NA1}\\
&&\{\hat{f},\ \hat{g}\}\in \{\hat{P}^x\hat{T},\ \hat{P}^y\hat{C},\ \hat{P}^x\hat{T}\hat{P}^y\hat{C}\},\ \hat{f}\neq \hat{g}.\nonumber
\end{eqnarray}
For the KPII system, we only write down two special Abelian real two-place nonlocal reductions from \eqref{NA},
\begin{eqnarray}
&&p_{xt}+3p_{yy}+[p_{xx}+6p^2 +3(p-p(-x,y,-t))^2]_{xx}=0,\label{F=1}
\end{eqnarray}
and
\begin{eqnarray}
&&p_{xt}+3p_{yy}+[p_{xx}+6p(-x,y,-t) (2p-p(-x,y,-t))]_{xx}=0.\label{G=1}
\end{eqnarray}

To end this section, we write down a general vector form of a special local and nonlocal KP system
\begin{eqnarray}
&&p_{xt}+3\sigma^2p_{yy}+[p_{xx}+6(P|U|P)]_{xx}=0,\label{GKP}\\
&& (P|U|P)\equiv \sum_{i, j=1}^4U_{ij}p_ip_j,\ U_{ij}=0,\ \forall i>j,\nonumber \\
&&p_1=p,\ p_2=p^{\hat{f}},\ p_3=p^{\hat{g}},\ p_4=p^{\hat{f}\hat{g}},\ f,\ g \in \{1,\ \hat{P}^x\hat{T},\ \hat{P}^y,\ \hat{P}^y\hat{P}^x\hat{T}\}. \nonumber
\end{eqnarray}
The model equation \eqref{GKP} is a generalization of examples given by \eqref{KPe1} and \eqref{KPe2}.

\section{Exact solutions of multi-place nonlocal KP systems}
\subsection{Symmetry-antisymmetry separation approach to solve nonlocal systems}
For a second order operator, $\hat{g}$,
\begin{equation}
 \hat{g}^2=1,\
\end{equation}
one can always separate an arbitrary function, $A$, as a summation of $\hat{g}$-symmetric and $\hat{g}$-antisymmetric parts in the following way,
\begin{eqnarray}
&&A=\frac12(A+A^{\hat{g}})+\frac12(A-A^{\hat{g}})\equiv u+v, \label{Auv}\\
&&u\equiv \frac12(A+A^{\hat{g}}),\ v\equiv \frac12(A-A^{\hat{g}}). \label{uv}
\end{eqnarray}
It is clear that $u$ and $v$ defined in \eqref{uv} are symmetric and anti-symmetric, respectively, with respect to $\hat{g}$, i.e.,
\begin{equation}
\hat{g} u=u,\ \hat{g} v=-v.\label{S-AS}
\end{equation}
Thus, a two-place nonlocal system
\begin{equation}
F(A,\ B)=0,\quad B=A^{\hat{g}},\ g^2=1,\label{F}
\end{equation}
can be transformed to a coupled
local system
\begin{eqnarray}
&&F_1(u,\ v)=0,\quad F_1=F+\hat{g}F,\label{F1}\\
&&F_2(u,\ v)=0,\quad F_2=F-\hat{g}F,\label{F2}
\end{eqnarray}
by using \eqref{Auv}.
Therefore, to solve the nonlocal equation \eqref{F} is equivalent to solving the local system \eqref{F1} and \eqref{F2} with \eqref{uv}.

Similarly, a four-place nonlocal system
\begin{equation}
F(p,\ q,\ r,\ s)=0,\quad q=p^{\hat{f}},\ r=p^{\hat{g}},\ s=p^{\hat{f}\hat{g}},\ \hat{f}^2=\hat{g}^2=1,\label{F4}
\end{equation}
can be changed to a coupled
local system
\begin{eqnarray}
&&F_1(u,\ v,\ w,\ z)=0,\quad F_1=F+\hat{g}F+\hat{f}F+\hat{f}\hat{g}F,\label{F41}\\
&&F_2(u,\ v,\ w,\ z)=0,\quad F_2=F+\hat{f}F-\hat{g}F-\hat{f}\hat{g}F,\label{F42}\\
&&F_3(u,\ v,\ w,\ z)=0,\quad F_3=F+\hat{g}F-\hat{f}F-\hat{f}\hat{g}F,\label{F43}\\
&&F_4(u,\ v,\ w,\ z)=0,\quad F_4=F+\hat{f}\hat{g}F-\hat{g}F-\hat{f}F,\label{F44}
\end{eqnarray}
by using the symmetric-antisymmetric separation
\begin{eqnarray}
p=u+v+w+z\label{Auvwz},
\end{eqnarray}
such that
\begin{eqnarray}
&&u=\frac14\big(p+p^{\hat{f}}+p^{\hat{g}} +p^{\hat{f}\hat{g}}\big),
v=\frac14\big(p+p^{\hat{f}}-p^{\hat{g}} -p^{\hat{f}\hat{g}}\big),\label{ruv}\\
&&w=\frac14\big(p+p^{\hat{g}}-p^{\hat{f}} -p^{\hat{f}\hat{g}}\big),\
z=\frac14\big(p+p^{\hat{f}\hat{g}}-p^{\hat{g}} -p^{\hat{f}}\big).\label{rwz}
\end{eqnarray}
From the definitions \eqref{ruv} and \eqref{rwz}, it is not difficult to find that $u$ is group
$${\cal{G}}=\{1,\ \hat{f},\ \hat{g},\ \hat{f}\hat{g}\}$$
invariant, $v$ is $\hat{f}$ invariant and $\hat{g}$ antisymmetric, $w$ is $\hat{g}$ invariant and $\hat{f}$ antisymmetric, while $z$ is both $\hat{f}$ and $\hat{g}$ antisymmetric. To sum up, we have
\begin{eqnarray}
&&\hat{f}u=\hat{g}u=\hat{f}\hat{g}u=u,\ \hat{f}v=-\hat{g}v=-\hat{f}\hat{g}v=v,\nonumber\\
&&
\hat{g}w=-\hat{f}w=-\hat{f}\hat{g}w=w,\
\hat{f}\hat{g}z=-\hat{g}z=-\hat{f}z=z. \label{uvwz1}
\end{eqnarray}
Hence, to solve the nonlocal equation \eqref{F4} is equivalent to solving the local system \eqref{F41}-\eqref{F44} with the conditions \eqref{uvwz1}.
\subsection{Exact multiple soliton solutions of a two-place nonlocal KP equation}
For concreteness, we study the exact solutions of the special two-place nonlocal KP equation
\begin{equation}
A_{xt}+A_{xxxx}+\frac32[(A+B)(3A+B)_x]_x+3\sigma^2A_{yy}=0,\ B=A^{\hat{g}},\ \hat{g}\in
\{\hat{P}^x\hat{T},\ \hat{P}^y,\ \hat{P}^y\hat{P}^x\hat{T}\}. \label{KP2}
\end{equation}
Using the symmetry-antisymmetry separation procedure,
\begin{equation}
A=u+v, \quad \hat{g}u=u,\quad \hat{g} v=-v.
\end{equation}
\eqref{KP2} is separated to
\begin{eqnarray}
&&u_{xt}+(u_{xx}+6u^2)_{xx}+3\sigma^2 u_{yy}=0,\label{KPu}\\
&&(v_{t}+v_{xxx}+6uv)_{x}+3\sigma^2 v_{yy}=0. \label{KPv}
\end{eqnarray}

The multiple soliton solutions of the KP equation \eqref{KPu} can be simply obtained by using the well known Hirots's bilinear approach. The bilinear form of \eqref{KPu} can be written as
\begin{eqnarray}
&&(D_xD_t+D_x^4+3\sigma^2D_y^2)\psi\cdot\psi=0,\label{BLKPu}
\end{eqnarray}
by means of the transformation
\begin{eqnarray}
&&u=(\ln \psi)_{xx},\label{Tru}
\end{eqnarray}
where the bilinear operators $D_x,\ D_t$ and $D_y$ are defined by
\begin{eqnarray*}
&&D_x^mD_t^nD_y^p f\cdot g=\left.(\partial_x-\partial_{x'})^m (\partial_t-\partial_{t'})^n (\partial_y-\partial_{y'})^p f(x,\ y,\ t)g(x',\ y',\ t')\right|_{x'=x,y'=y,t'=t}.
\end{eqnarray*}
It is interesting that for the equation \eqref{KPv} with \eqref{Tru}, we have a special solution
\begin{eqnarray}
&&v=a(\ln \psi)_x\label{Trv}
\end{eqnarray}
with $a$ being an arbitrary constant.

Though $\{\eqref{Tru},\ \eqref{Trv}\}$ solves \eqref{KPu} and \eqref{KPv}, however, to get the solution of the two-place nonlocal KP equation \eqref{KP2}, we have to check the nonlocal conditions \eqref{S-AS} for $\hat{g}=\hat{P}^y\hat{P}^x\hat{T},\ \hat{P}^x\hat{T}$ and $\hat{P}^y$, respectively.

Case 1. $\hat{g}=\hat{P}^y\hat{P}^x\hat{T}$. In this case, the multi-soliton solutions of the two-place KP equation \eqref{KP2} can be written as
\begin{eqnarray}
&&A=u+v=\left(\partial_x^2+a\partial_x\right)\ln \psi,\quad \psi=\sum_{\{\nu\}}K_{\{\nu\}}
\cosh\left(\frac12\sum_{j=1}^N\nu_j\eta_j\right),\label{PPT}\\
&&K_{\{\nu\}}=\prod_{i>j}^N\sqrt{3k_i^2k_j^2(k_i-\nu_i\nu_jk_j)^2 -\sigma^2(l_ik_j-l_jk_i)^2},\nonumber\\
&&\eta_j=k_jx+l_jy-(k_j^3+\sigma^2k_j^{-1}l_j^2)t,\nonumber
\end{eqnarray}
where the summation on $\{\nu\}\equiv \{\nu_1,\ \nu_2,\ \ldots,\ \, \nu_i,\ \ldots,\ \nu_N\}$ should be done for all possible permutations $\nu_i=\{1,\ -1\},\ i=1,\ 2,\ \ldots,\ N$.

Case 2. $\hat{g}=\hat{P}^x\hat{T}$. In this case, the multiple soliton solution of the two-place nonlocal KP equation \eqref{KP2} still possesses the form \eqref{PPT}. However, the paired condition has to be satisfied,
\begin{eqnarray}
&&N=2n,\ k_{n+i}=\pm k_i,\ l_{n+i}=\mp l_i.\label{PT}
\end{eqnarray}
The condition \eqref{PT} implies that the odd numbers of soliton solutions in the form \eqref{PPT} are prohibited for the partially inverse nonlocal system KP system \eqref{KP2} with $\hat{g}=\hat{P}^x\hat{T}$. Under the condition \eqref{PT}, we have paired travelling wave variables \begin{equation}\eta\equiv\{\eta_i=k_ix+l_iy-(k_j^3+\sigma^2k_j^{-1}l_j^2)t,\ \eta_{n+i}=\pm k_ix+\mp l_iy\mp (k_j^3+\sigma^2k_j^{-1}l_j^2)t,\ ,i=1,\ 2,\ \ldots,\ n\}
\end{equation}
with the property
\begin{equation}
\hat{P}^x\hat{T}\eta=\mp\eta.
\end{equation}
Thus, the nonlocal condition \eqref{S-AS} is naturally satisfied for  $\hat{g}=\hat{P}^x\hat{T}$.

For $n=1\ (N=2)$, the solution \eqref{PPT} with \eqref{PT} becomes
\begin{eqnarray}
&&A=(\partial_x^2+a\partial_x)\ln F_2,\\
&&F_2=2k_1l_1\sigma\cosh(2l_1y)+2k_1\sqrt{\sigma^2l_1^2-4k_1^4} \cosh\big[2(k_1x-4k_1^3t-3k_1^{-1}l_1^2\sigma^2t)\big].\label{F21}
\end{eqnarray}

For $n=2\ (N=4)$, the solution \eqref{PPT} with \eqref{PT} possess the form
\begin{eqnarray}
&&A=(\partial_x^2+a\partial_x)\ln F_4,\\
&&F_4=K_{\{1,1,1,-1\}}[\cosh(\xi)+\cosh(\hat{g}\xi)]
+K_{\{1,1,-1,1\}}[ \cosh(\eta)+\cosh(\hat{g}\eta)]\nonumber\\
&&\qquad
+K_{\{1,1,1,1\}}\cosh(\tau)+K_{\{1,-1,1,-1\}}\cosh(\tau_1) \nonumber\\
&&\qquad+K_{\{1,1,-1,-1\}}\cosh[2(l_1+l_2)y] +K_{\{1,-1,-1,1\}}\cosh[2(l_1-l_2)y],\label{rF4}\\
&&\xi=2k_1x+2l_2y-2(4k_1^3+3k_1^{-1}l_1^2\sigma^2)t,\
\eta=2k_2x+2l_1y-2(4k_2^3+3k_2^{-1}l_2^2\sigma^2)t,\nonumber\\
&&\tau=2(k_1+k_2)x-2(4k_1^3+3\sigma^2 k_1^{-1} l_1^2)t-2(4k_2^3+3\sigma^2 k_2^{-1}l_2^2)t.\nonumber
\end{eqnarray}

Case 3. $\hat{g}=\hat{P}^y$. In this case, the multiple soliton solution form \eqref{PPT} is correct only for the conditions \eqref{PT} and
\begin{equation}
a=0
\end{equation}
being satisfied for the two-place nonlocal KP equation with $\hat{g}=\hat{P}^y$. In other words, for the third kind of two-place nonlocal KP equation \eqref{KP2}, we have not yet found $\hat{P}^y$-symmetry breaking multiple soliton solutions.

\subsection{Exact multiple soliton solutions of a four-place nonlocal KP equation}
In this subsection, we study the possible multiple soliton solutions for the four-place nonlocal KP equation \eqref{GKP}.
By using the symmetric-antisymmetric separation relations \eqref{Auvwz}, \eqref{ruv} and \eqref{rwz}, the four-place nonlocal KP equation can be equivalent to
\begin{eqnarray}
&&u_{xt}+(u_{xx}+c_+u^2+c_-z^2+e_+v^2+e_-w^2)_{xx} +3\sigma^2u_{yy}=0,\label{KPU}\\
&&v_{xt}+(v_{xx}+d_{+}uv+d_{-}wz)_{xx} +3\sigma^2v_{yy}=0,\label{KPV}\\
&&w_{xt}+(w_{xx}+b_{+}uw+b_{-}vz)_{xx} +3\sigma^2w_{yy}=0,\label{KPW}\\
&&z_{xt}+(z_{xx}+a_{+}wv+a_{-}uz)_{xx} +3\sigma^2z_{yy}=0\label{KPZ}
\end{eqnarray}
with the symmetric-antisymmetric conditions \eqref{uvwz1} and the constant relations
\begin{eqnarray*}
&&c_{\pm}=U_{11}\pm U_{12}\pm U_{13}+U_{14}+U_{22} +U_{23}\pm U_{24}+U_{33}\pm U_{34}+U_{44},\\
&&e_{\pm}=U_{11}\mp U_{12}\pm U_{13}-U_{14}+U_{22} -U_{23}\pm U_{24}+U_{33}\mp U_{34}+U_{44},
\end{eqnarray*}
\begin{eqnarray*}
&&d_{\pm}=2(U_{11}\pm U_{13}-U_{22}\mp U_{24}+U_{33}-U_{44}),\\
&&b_{\pm}=2(U_{11}\pm U_{12}+U_{22}-U_{33}\mp U_{34}-U_{44}),\\
&&a_{\pm}=2(U_{11}\mp U_{14}-U_{22}\pm U_{23}-U_{33}+U_{44}).
\end{eqnarray*}
The system of equations \eqref{KPU}-\eqref{KPZ} is not integrable for arbitrary constants $\{a_{\pm},\ b_{\pm},\ c_{\pm},\ d_{\pm},\ e_{\pm}\}$. For some special fixed parameters, for instance,
\begin{eqnarray}
c_+=3,\ c_- = e_+= e_-=d_-=b_-= 0,\ d_+=b_+=a_+=a_-=6,\label{cde}
\end{eqnarray}
the four-place nonlocal equation \eqref{GKP} becomes
\begin{eqnarray}
&&p_{xt}+3\sigma^2p_{yy}+\left[p_{xx} -3u^2 +6pu +\frac38(p-p^{\hat{f}\hat{g}})^2 -\frac38(p^{\hat{f}}-p^{\hat{g}})^2\right]_{xx}=0,\label{GKPA}\\
&&u\equiv \frac14(p+p^{\hat{f}}+p^{\hat{g}}+p^{\hat{f}\hat{g}}),\quad \hat{f}\equiv\hat{P}^x\hat{T},\ \hat{g}\equiv\hat{P}^y, \nonumber
\end{eqnarray}
while the related symmetric-antisymmetric system \eqref{KPU}-\eqref{KPZ} becomes an integrable coupling  system
\begin{eqnarray}
&&u_{xt}+(u_{xx}+3u^2)_{xx}+3\sigma^2u_{yy}=0,\label{KU}\\
&&v_{xt}+(v_{xx}+6uv)_{xx}+3\sigma^2v_{yy}=0,\label{KV}\\
&&w_{xt}+(w_{xx}+6uw)_{xx}+3\sigma^2w_{yy}=0,\label{KW}\\
&&z_{xt}+(z_{xx}+6wv+6uz)_{xx}+3\sigma^2z_{yy}=0.\label{KZ}
\end{eqnarray}
Because \eqref{KV} and \eqref{KW} is just the symmetry equations of \eqref{KU} and the system of equations \eqref{KW} and \eqref{KZ} is also a symmetry system of \eqref{KU} and \eqref{KV}, it is not difficult to find some special solutions of \eqref{KU}-\eqref{KZ} and then the solutions of the four-place nonlocal KP equation \eqref{GKPA}. A special multiple soliton solutions of \eqref{GKPA} can be written as
\begin{equation}
p=2(1+\beta_1\partial_y+\beta_2\partial_x +\beta_1\beta_2\partial_x\partial_y)(\ln \psi)_{xx},
\end{equation}
where $\psi$ is given in \eqref{PPT} with the paired condition \eqref{PT} satisfying the symmetric-antisymmetric conditions \eqref{uvwz1}. $\psi=F_2$ with \eqref{F21} and $\psi=F_4$ with \eqref{rF4} are two simplest two-soliton and four-soliton examples.

\section{Summary and discussions}
In summary, the two-place nonlocal integrable models are systematically
 extended to multi-place nonlocal integrable (and nonintegrable) nonlinear models by means of the discrete symmetry reductions of the coupled local systems. Especially, various four-place nonlocal integrable systems are obtained.

Starting from every multi-component AKNS system, one may derive some local and nonlocal multi-place AKNS, NLS and Melnikov systems. For instance, from the two-component AKNS system \eqref{CAKNSa}, one can obtain the usual local AKNS system \eqref{ABAKNS2} with $\hat{f}=1$, local NLS equation \eqref{ABAKNS2} with $\{\hat{f}=1,\ r=q^*\}$, local Melnikov system \eqref{ABAKNS1a} with $\hat{g}=\hat{C}$, three types of two-place nonlocal AKNS systems \eqref{ABAKNS2} with $\hat{f} \neq 1$, three types of two-place nonlocal Melnikov models \eqref{ABAKNS1a} with $\hat{g} \neq \hat{C}$, nine types of two-place nonlocal NLS equations \eqref{NLSV} with \eqref{V1}-\eqref{V6} and $\{\alpha,\ \beta\}=\{\alpha,\ 0\}$ or $\{\alpha,\ \beta\}=\{0,\ \alpha\}$, and six types of four-place nonlocal NLS systems \eqref{NLSV} with \eqref{V7}-\eqref{V8} and $\{\alpha,\ \beta\}=\{\alpha,\ 0\}$ or $\{\alpha,\ \beta\}=\{0,\ \alpha\}$.

In fact, starting from every coupled nonlinear systems, one may also find some types of multi-place nonlocal systems via discrete symmetry groups. In addition to the NLS equation, the (2+1)-dimensional KP equation is another important physically applicable model. To find some types of multi-place extensions of the KP equation, the matrix KP equations are best candidates. In this paper, some types of multi-place nonlocal KP equations are obtained from the $\hat{P}\hat{T}\hat{C}$ symmetry reductions from some special Abelian and non-Abelian matrix KP equations.

Because many nonlocal nonlinear systems can be derived from the $\hat{P}\hat{T}\hat{C}$ symmetry reductions, the nonlocal systems may be solved via $\hat{P}$-$\hat{T}$-$\hat{C}$ symmetric-antisymmetric separation approach (SASA). Using SASA, the two-place nonlocal KP equation \eqref{KP2} and four-place nonlocal KP equation \eqref{GKPA} are explicitly solved for special types of multiple soliton solutions.

\section*{Acknowledgements}
The author is grateful  to thank Professors X. Y. Tang, D. J. Zhang, Z. N. Zhu, Q. P. Liu, X. B. Hu, Y. Q. Li and Y. Chen for their helpful discussions. The work was sponsored by the Global Change Research
Program of China (No.2015CB953904),  Shanghai Knowledge Service Platform for Trustworthy Internet of Things (No. ZF1213), the National Natural Science Foundations of China (No. 11435005) and K. C. Wong Magna Fund in Ningbo University.

\end{document}